\documentclass{article}
\newcommand{\be}{\begin{equation}}
\newcommand{\ee}{\end{equation}}
\newcommand{\nbar}[1]{\overline{#1}}

\def\bea{\begin{eqnarray}}
\def\eea{\end{eqnarray}}
\def\beas{\begin{eqnarray*}}
\def\eeas{\end{eqnarray*}}
\def\sla{\raise.15ex\hbox{$/$}\kern-.57em}

\def\parm{{\partial}^{-}}
\def\parp{\partial^+}

\usepackage{amsmath}
\usepackage{latexsym}
\usepackage{graphicx}

\begin{document}
\begin{titlepage}
\begin{flushright}    AEI-2005-154 \\ 
\end{flushright}
\vskip 1cm
\centerline{\LARGE {\bf {Theories with Memory}}}

\vskip 2cm

\centerline{Sudarshan Ananth} 
\vskip .5cm
\centerline{\em  Max-Planck-Institut f\"{u}r Gravitationsphysik}
\centerline{\em Albert-Einstein-Institut, D-14476 Golm}
\vskip 1.5cm

\vskip 1cm

\centerline{\bf {Abstract}}

\vskip .5cm

\noindent Dimensionally reduced supersymmetric theories retain a great deal of information regarding their higher dimensional origins. In superspace, this ``memory" allows us to restore the action governing a reduced theory to that describing its higher-dimensional progenitor. We illustrate this by restoring four-dimensional $\mathcal N=4$ Yang-Mills to its six-dimensional parent, $\mathcal N=(1,1)$ Yang-Mills. Supersymmetric truncation is introduced into this framework and used to obtain the $\mathcal N=1$ action in six dimensions. We work in light-cone superspace, dealing exclusively with physical degrees of freedom.
\end{titlepage}

\section{Introduction}

\noindent Dimensionally reduced supersymmetric theories retain a great deal of information regarding their higher dimensional origins. $(\mathcal N=4,d=4)$ SuperYang-Mills is a good example of a theory with such ``memory". The six scalars in its spectrum serve as signatures of a lost (compactified) $SO(6)$ while its $SU(4)$ spinors assemble nicely into a single eight-spinor: ${\bf 4}_{1/2}+{\bf \bar4}_{-1/2}~=~{\bf 8}_s$ of $SO(8)$. Its spectrum naturally favors reformulation in ten dimensions with a single supersymmetry. 

\vskip 0.3cm

\noindent In superspace, this allows us to ``oxidize" (restore) four-dimensional $\mathcal N=4$ Yang-Mills into its fully ten-dimensional parent, $\mathcal N=1$ Yang-Mills. This is achieved by simply generalizing the $d=4$ transverse space derivatives~\cite{ABR1}.

\vskip 0.3cm

\noindent The six scalars in the $\mathcal N=4$ spectrum could equally well be thought of as $2\,+\,4$ scalars with the first two signaling a six-dimensional progenitor. Indeed, we will show that the $\mathcal N=4$ action can be restored to that describing six-dimensional $\mathcal N=(1,1)$ Yang-Mills. Supersymmetric truncation is introduced into this framework and used (as illustrated below) to obtain the $(\mathcal N=1,d=6)$ action.

\hskip 0.8cm \includegraphics[scale=0.37]{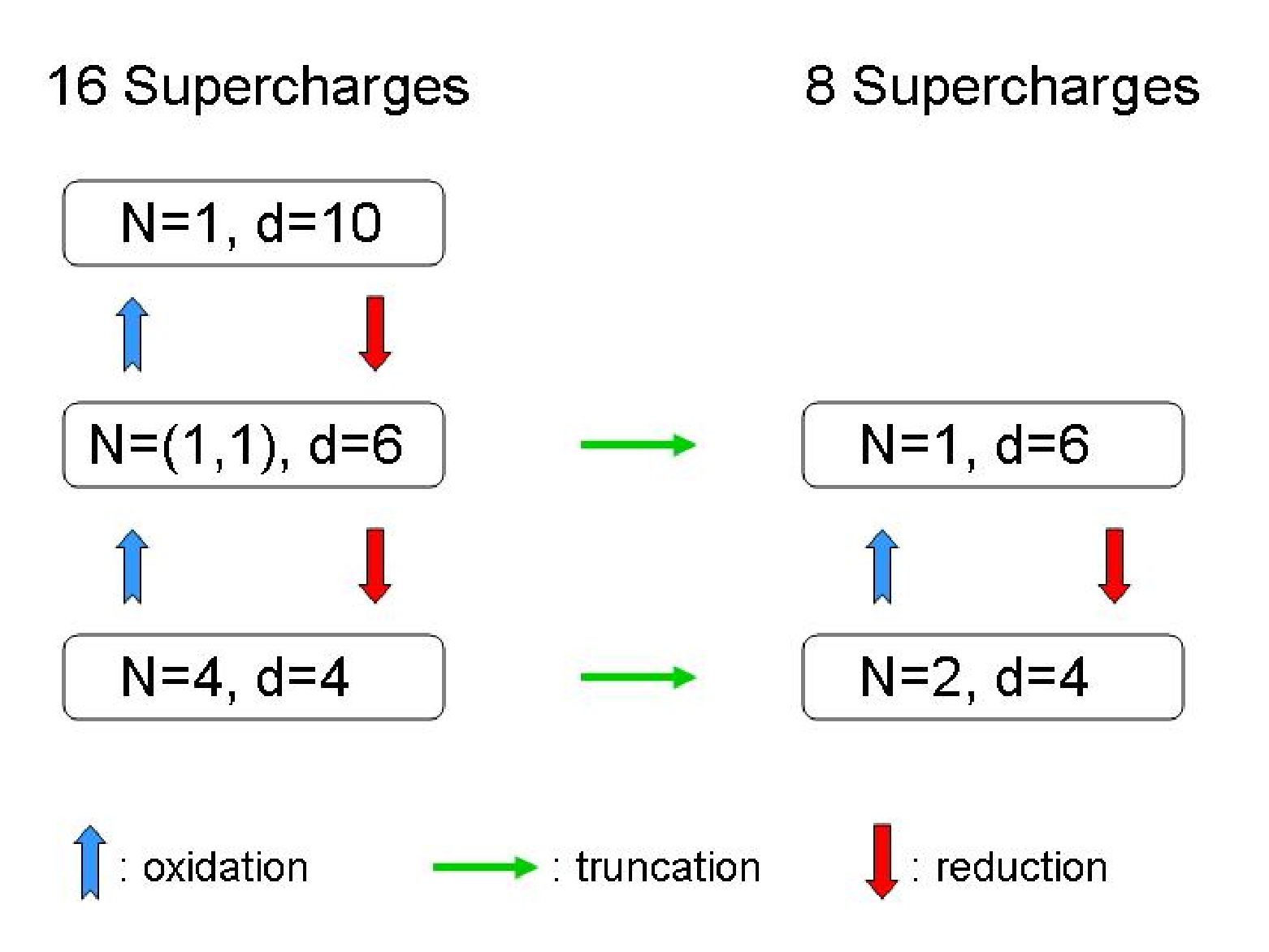}

\noindent In six dimensions, massless particles are classified according to the little group $SO(4)$. Our focus (in this paper) will be mainly on the $\mathcal N=(1,1)$ Yang-Mills theory in six dimensions. This theory has $16$ supercharges and may be obtained by reduction from ten-dimensional $\mathcal N=1$ Yang-Mills. The relevant little group decomposition is
\beas
SO(8)\,\supset~SO(4)~\times~SO(4)\ .
\eeas
\noindent The first $SO(4)\,\sim\,SU(2)\,\times\,SU(2)$ is the little group in six dimensions while the second represents the R-symmetry in the theory. The bosonic fields in the $\mathcal N=(1,1)$ theory include a gauge field and four scalars. Under the little group, its spectrum transforms as
\beas
(\,{\bf {2}},\,{\bf {2}}\,)\,+\,4\cdot(\,{\bf {1}},\,{\bf {1}}\,)\,+\,2\cdot(\,{\bf {2}},\,{\bf {1}}\,)\,+\,2\cdot(\,{\bf {1}},\,{\bf {2}}\,)\ .
\eeas

\vskip 0.3cm

\noindent There is a second theory in six dimensions with $16$ supercharges. This is the superconformally-invariant $\mathcal N=(2,0)$ theory characterized by an $SO(5)$ R symmetry. The spectrum of this theory transforms as
\beas
(\,{\bf {3}}\,,\,1\,)\,+\,5\cdot(\,{\bf {1}},\,{\bf {1}}\,)\,+\,4\cdot(\,{\bf {2}},\,{\bf {1}}\,)\ ,
\eeas
\noindent with a bosonic spectrum comprised of a self-dual antisymmetric tensor and five scalars. One of the aims of the present paper is to construct the six-dimensional SuperPoincar\'e algebra with a view to tackling this theory in the future. In reference~\cite{ABR3}, we developed an algorithm to construct the entire $(\mathcal N=4,d=4)$ action starting from a single Ansatz for its dynamical supersymmetry generator. This was possible thanks to the vast $PSU(2,2|4)$ symmetry in that theory. The hope is that a similar approach will offer insights into the structure of other symmetry-laden theories like the superconformal ones in three~\cite{NS1,AOY,JHS} and six~\cite{ABS,PSH,AHH} dimensions.

\vskip 1cm

\subsection{Brief Summary of the Proposal}

\vskip 0.3cm

\noindent In Section {\bf {2}}, we review the formulation of $(\mathcal N=4,d=4)$ SuperYang-Mills in light-cone superspace. It will turn out that the entire action can be written in terms of a single superfield as
\beas
\int d^4x\,\int\,d^4\theta\,d^4{\bar \theta}\,{\cal L}\ ,
\eeas
where
\beas
{\cal L}&=&-\bar\phi\,\frac{\Box_4}{\partial^{+2}}\,\phi
~+\frac{4g}{3}\,f^{abc}_{}\,\Big(\,\frac{1}{\parp}\,{\bar \phi}^a\,\phi^b\,{\bar \partial}\,\phi^c\,+\,\frac{1}{\parp}\,\phi^a\,{\nbar \phi}^b\,\partial\,{\nbar \phi}^c\,\Big)\cr
&&-g^2f^{abc}_{}\,f^{ade}_{}\Big(\,\frac{1}{\partial^+_{}}(\phi^b\,\partial^+\phi^c)\frac{1}{\partial^+_{}}\,(\bar \phi^d_{}\,\partial^+_{}\,\bar\phi^e)+\frac{1}{2}\,\phi^b_{}\bar\phi^c\,\phi^d_{}\,\bar\phi^e\Big)\ .
\eeas
\vskip 0.3cm

\noindent The basic idea is to modify the above action such that it describes the six-dimensional $\mathcal N=(1,1)$ theory. This modification (sections {\bf {3}} and {\bf {4}}) is done in three steps;

$\\ \bullet\;$ Introduce two new coordinates (and their derivatives)

$\\ \bullet\;$ Allow $\phi$ and $\nbar \phi$ to depend on these new coordinates

$\\ \bullet\;$ Generalize the $d=4$ transverse derivatives $\partial$ and $\bar\partial$ to incorporate the new derivatives.

\vskip 0.5cm

\noindent The new ``generalized" derivatives ($\nabla$ and $\nbar \nabla$) are defined in section {\bf {4.1}}. Thus our proposal for the action governing the $d=6$ $\mathcal N=(1,1)$ theory is

\beas
\label{n=11}
\int d^6x\int d^4\theta\,d^4 \bar \theta\,{\cal L}\ ,
\eeas
where
\beas
{\cal L}&=&-\bar\phi\,\frac{\Box_6}{\partial^{+2}}\,\phi
~+\frac{4g}{3}\,f^{abc}_{}\,\Big(\frac{1}{\partial^+_{}}\,\bar\phi^a_{}\,\phi^b_{}\,{\nbar \nabla}\,\phi^c_{}+\frac{1}{\partial^+_{}}\,\phi^a_{}\,\bar\phi^b_{}\,\nabla\,\bar\phi^c_{}\Big)\cr
&&-g^2f^{abc}_{}\,f^{ade}_{}\Big(\,\frac{1}{\partial^+_{}}(\phi^b\,\partial^+\phi^c)\frac{1}{\partial^+_{}}\,(\bar \phi^d_{}\,\partial^+_{}\,\bar\phi^e)+\frac{1}{2}\,\phi^b_{}\bar\phi^c\,\phi^d_{}\,\bar\phi^e\Big)\ .
\eeas

\vskip 0.3cm

\noindent We will prove that this action is invariant under the six-dimensional SuperPoincar\'e algebra. This proof is presented in sections {\bf {4.1}} and {\bf {4.2}}.

\vskip 0.2cm

\noindent Finally, in section {\bf {5}} this six-dimensional $\mathcal N=(1,1)$ theory (with $16$ supercharges) is truncated to obtain the $(\mathcal N=1,d=6)$ theory with $8$ supercharges. Supersymmetric truncation~\cite{BT} is based on the fact that
\beas
\label{trunc1}
\int\,{d^6}x\,{d^{\,\mathcal N}}\theta\,{d^{\,\mathcal N}}{\bar \theta}\,{\cal L}\;\propto\;\int\,{d^6}x\,{d^{\,\mathcal N-1}}\theta\,{d^{\,\mathcal N-1}}{\bar \theta}\,{\bar d}_{\,\mathcal N}\,d^{\,\mathcal N}\,{\cal L}\,|_{\theta^{\,\mathcal N}\,=\,{\bar \theta}_{\,\mathcal N}\,=0}\ .
\eeas

\vskip 0.5cm

\noindent Six-dimensional $\mathcal N=1$ Yang-Mills has been studied previously in harmonic superspace by Howe, Stelle and West~\cite{HSW}. The theory was formulated in terms of $d=6$ superfields by Tollsten~\cite{AT} and Howe, Sierra and Townsend~\cite{HST}. The free six-dimensional hypermultiplet was formulated in terms of four-dimensional $\mathcal N=1$ superspace in references~\cite{WS,GPM}.

\vskip 1cm

\section{${\mathcal N}=4$ Yang-Mills in Light-Cone Superspace}

\vskip 0.3cm

\noindent With the space-time metric $(-,+,+,+)$, the light-cone coordinates and their derivatives are 
\bea
\begin{split}
{x^{\pm}}&=&\frac{1}{\sqrt 2}\,(\,{x^0}\,{\pm}\,{x^3}\,)\ ;\qquad ~ {\partial^{\pm}}=\frac{1}{\sqrt 2}\,(\,-\,{\partial_0}\,{\pm}\,{\partial_3}\,)\ , \\
x &=&\frac{1}{\sqrt 2}\,(\,{x_1}\,+\,i\,{x_2}\,)\ ;\qquad  {\bar\partial} =\frac{1}{\sqrt 2}\,(\,{\partial_1}\,-\,i\,{\partial_2}\,)\ , \\
{\bar x}& =&\frac{1}{\sqrt 2}\,(\,{x_1}\,-\,i\,{x_2}\,)\ ;\qquad  {\partial} =\frac{1}{\sqrt 2}\,(\,{\partial_1}\,+\,i\,{\partial_2}\,)\ .
\end{split}
\eea

\noindent We introduce $SU(4){\,\sim\,}SO(6)$ spinors $\theta^m$ and their conjugates ${\bar \theta}_n$ ($m\,,\,n\,=\,1,2,3,4$). The $d=4$ d'Alembertian reads
\bea
\Box_4\,=\,2\,(\,\partial\,{\bar \partial}\,-\,\parp\,\parm\,)\ .
\eea

\vskip 0.3cm

\noindent All the physical degrees of freedom of the $\mathcal N=4$ theory are captured by a single complex superfield~\cite{BLN1,BLN2}
\bea
\begin{split}
\phi\,(y)&=&\frac{1}{\parp}\,A\,(y)\,+\,\frac{i}{\sqrt 2}\,\theta^m\,\theta^n\,{\nbar C}_{mn}\,(y)\,+\,\frac{1}{12}\,\theta^m\,\theta^n\,\theta^p\,\theta^q\,\epsilon_{mnpq}\,\parp\,{\bar A}\,(y)\cr
&&~+~\frac{i}{\parp}\,\theta^m\,{\bar \chi}_m\,(y)\,+\,\frac{\sqrt 2}{6}\,\theta^m\,\theta^n\,\theta^p\,\epsilon_{mnpq}\,\chi^q\,(y) \ ,
\end{split}
\eea
\noindent with the $\frac{1}{\parp}$ interpreted as~\cite{KS}
\bea
\frac{1}{\parp}\,f\,(\,x^-\,)\,=\,\frac{1}{2}\,\int\,\epsilon\,(\,\xi\,-\,x^-\,)\;f\,(\,\xi\,)\;d\xi\ .
\eea
\noindent The gauge fields appear as
\be
A\,=\,\frac{1}{\sqrt 2}\,(A_1+i\,A_2)\ ;\qquad {\bar A}\,=\,\frac{1}{\sqrt 2}\,(A_1-i\,A_2)\ ,
\ee
\noindent the six scalars as $SU(4)$ bispinors
\bea
\label{bisp}
C^{m\,4}\,=\,\frac{1}{\sqrt 2}\,({A_{m+3}}\,+\,i\,{A_{m+6}})\ ;\qquad {\nbar C}_{mn}\,=\,\frac{1}{2}\,\epsilon_{mnpq}\,C^{pq}\ ,
\eea
\noindent (for $m\;\neq\,4$) and the fermi fields as $\chi^m$ and ${\bar \chi}_n$. All fields are local in the modified light-cone coordinates
\bea
y\,=\,(\,x,\,{\bar x},\,{x^+},\,y^-\,\equiv\,x^-\,-\,\frac{i}{\sqrt 2}\,\theta^m\,{\bar \theta}_m\,)\ .
\eea

\noindent The $\mathcal N=4$ Yang-Mills light-cone action is then simply
\be
\label{n=4}
72\,\int\,d^4x\int d^4\theta\,d^4 \bar \theta\,{\cal L}\ ,
\ee
where
\bea
{\cal L}&=&-\bar\phi\,\frac{\Box_4}{\partial^{+2}}\,\phi
~+\frac{4g}{3}\,f^{abc}_{}\,\Big(\,\frac{1}{\parp}\,{\bar \phi}^a\,\phi^b\,{\bar \partial}\,\phi^c\,+\,\frac{1}{\parp}\,\phi^a\,{\nbar \phi}^b\,\partial\,{\nbar \phi}^c\,\Big)\cr
&&-g^2f^{abc}_{}\,f^{ade}_{}\Big(\,\frac{1}{\partial^+_{}}(\phi^b\,\partial^+\phi^c)\frac{1}{\partial^+_{}}\,(\bar \phi^d_{}\,\partial^+_{}\,\bar\phi^e)+\frac{1}{2}\,\phi^b_{}\bar\phi^c\,\phi^d_{}\,\bar\phi^e\Big)\ .
\eea
where the $f^{abc}$ are the structure functions of the Lie algebra and Grassmann integration is normalized so that $\int d^4\theta\;\theta^1\theta^2\theta^3\theta^4=1$.

\vskip 1cm

\subsection{The $d=4$ SuperPoincar\'e Algebra}

\vskip 0.3cm

\noindent The action in equation (\ref {n=4}) is left invariant by the $d=4$ SuperPoincar\'e algebra. We will simply write down the generators here and refer the reader to references~\cite{BLN1,BLN2,ABR2} for further details. 

\vskip 0.3cm

\noindent The bosonic generators include the four-momenta

\bea
p^+_{}\,&=\,-i\,\partial^+_{}\ ,\qquad p~=~-i\,\partial\ ,\qquad \bar p~=~-i\,\bar\partial\ ,\qquad p^-_{}~=~-i\frac{\partial\bar\partial}{\partial^+_{}}\ ,
\eea

\noindent the rotations

\bea
\begin{split}
\label{d=4g}
&j\,=\,x\,\bar\partial-\bar x\,\partial+\frac{1}{ 2}\,(\,\theta^m\,{\bar \partial}_m\,-\,{\bar \theta}_m\,\partial^m\,)\,+\frac{i}{4\sqrt{2}\,\parp}\,(\,d^m\,{\bar d}_m\,-\,{\bar d}_m\,d^m\,)\  \\\\
&j^+\,=\,i\, x\,\partial^+_{}\ ,\qquad \qquad \bar j^+\,=\,i\,\bar x\,\partial^+_{}\ , \\
&j^{+-}\,=\,i\,x^-_{}\,\parp-\frac{i}{2}\,(\,\theta^m\,{\bar \partial}_m\,+\,{\bar \theta}_m\,\partial^m\,)\ ,  \\\\
\end{split}
\eea

\noindent and the boosts
\bea
\begin{split}
&j^-\,=i\,x\,\frac{\partial{\bar \partial}}{\parp}\,-\,i\,x^-_{}\,\partial~+~i\,\Big( \theta^m\,{\bar \partial}_m\,+\,\frac{i}{4\sqrt{2}\,\partial^+}\,(\,d^m\,{\bar d}_m\,-\,{\bar d}_m\,d^m\,)\Big)\frac{\partial}{\parp}\ , \\ 
&\bar j^-\,=i\,\bar x\,\frac{\partial\bar\partial}{\parp}~-~i\,x^-_{}\,{\bar \partial}~+~ i\,\Big({\bar \theta}_n\,\partial^n\,+\,\frac{i}{4\sqrt{2}\,\parp}\,(\,d^n\,{\bar d}_n\,-\,{\bar d}_n\,d^n\,)\,\Big)\frac{\bar\partial}{\parp}\ .
\end{split}
\eea

\vskip 0.3cm

\noindent The fermionic operators are the chiral derivatives
\bea
{d^{\,m}}\,=\,-\,{\partial^m}\,-\,\frac{i}{\sqrt 2}\,{\theta^m}\,{\partial^+}\ ;\qquad{{\bar d}_{\,n}}=\;\;\;{{\bar \partial}_n}\,+\,\frac{i}{\sqrt 2}\,{{\bar \theta}_n}\,{\partial^+}\ ,
\eea
\noindent the kinematical supersymmetries
\bea
q^m_{\,+}\,=\,-\,\partial^m\,+\,\frac{i}{\sqrt 2}\,\theta^m\,\parp\ ;\qquad {\bar q}_{\,+\,n}=\;\;\;{\bar \partial}_n\,-\,\frac{i}{\sqrt 2}\,{\bar \theta}_n\,\parp\ ,
\eea
\noindent and the dynamical supersymmetries
\bea
q^m_{\,-}\,\equiv~i\,[\,\bar j^-\,,\,q^m_{\,+}\,]\,=\,\frac{\partial}{\parp}\,q^m_{\,+}\ ;\qquad 
{\bar q}_{\,-\,n}\,\equiv\,i\,[\,j^-\,,\,{\bar q}_{\,+\,n}\,]\,=\,\frac{\bar \partial}{\parp}\,{\bar q}_{\,+\,n}\ .
\eea

\vskip 0.3cm

\noindent The superfield and its complex conjugate satisfy chiral constraints,

\be
{d^{\,m}}\,\phi\,=\,0\ ;\qquad {\bar d_{\,m}}\,\bar\phi\,=\,0\ ,
\ee

\noindent as well as ``inside-out" constraints

\be
\label{ioc}
\bar d_m^{}\,\bar d_n^{}\,\phi~=~\frac{1}{ 2}\,\epsilon_{mnpq}^{}\,d^p_{}\,d^q_{}\,\bar\phi\ .
\ee

\vskip 0.3cm

\noindent The next step is to enlarge this SuperPoincar\'e algebra to six dimensions.

\vskip 1cm

\section{Six Dimensions}

\vskip 0.3cm

\noindent The reduction from six to four dimensions involves the little-group decomposition
\bea
SO(4)\,\supset~SO(2)~\times~SO(2)\ .
\eea

\noindent The first $SO(2)$ is described by the first generator in equation (\ref{d=4g}). In order to build the entire $d=6$ algebra, we need to introduce the second $SO(2)$ and the generators of the coset: $SO(4)/(SO(2){\times}SO(2))$.

\vskip 0.3cm

\noindent In the bispinor language of equation (\ref{bisp}) we may introduce upto six new coordinates as
\bea
x^{mn}\ ;\qquad {\bar x}_{mn}\,=\,\frac{1}{2}\,\epsilon_{mnpq}\,x^{pq}\ .
\eea
However, since we require only two additional directions, we choose to introduce only
\bea
\begin{split}
x^{12}\,&=\,\frac{1}{\sqrt 2}\,(\,x_6\,-\,i\,x_9\,)\ ;\qquad {\bar x}_{12}\,=\,\frac{1}{\sqrt 2}\,(\,x_6\,+\,i\,x_9\,)\ ,\\
x^{34}\,&=\,\frac{1}{\sqrt 2}\,(\,x_6\,+\,i\,x_9\,)\ ;\qquad {\bar x}_{34}\,=\,\frac{1}{\sqrt 2}\,(\,x_6\,-\,i\,x_9\,)\ ,
\end{split}
\eea

\noindent and their derivatives

\bea
\begin{split}
\partial^{12}\,&=\,\frac{1}{\sqrt 2}\,(\,\partial_6\,-\,i\,\partial_9\,)\ ;\qquad {\bar \partial}_{12}\,=\,\frac{1}{\sqrt 2}\,(\,\partial_6\,+\,i\,\partial_9\,)\ ,\\
\partial^{34}\,&=\,\frac{1}{\sqrt 2}\,(\,\partial_6\,+\,i\,\partial_9\,)\ ;\qquad {\bar \partial}_{34}\,=\,\frac{1}{\sqrt 2}\,(\,\partial_6\,-\,i\,\partial_9\,)\ .
\end{split}
\eea

\noindent The corresponding (new) $SO(2)$ generator is

\bea
\begin{split}
\cal J\,=\,&\frac{1}{2}\,(\,x^{12}\,{\bar \partial}_{12}\,-\,{\bar x}_{12}\,\partial^{12}\,)\,-\,\frac{1}{2}\,(\,x^{34}\,{\bar \partial}_{34}\,-\,{\bar x}_{34}\,\partial^{34}\,)\,+\,\frac{1}{2}\,(\,\theta^1\,{\bar \partial}_1\,+\,\theta^2\,{\bar \partial}_2\,-\,{\bar \theta}_1\,\partial^1\,-\,{\bar \theta}_2\,\partial^2\,) \\
&-\,\frac{1}{2}\,(\,\theta^3\,{\bar \partial}_3\,+\,\theta^4\,{\bar \partial}_4\,-\,{\bar \theta}_3\,\partial^3\,-\,{\bar \theta}_4\,\partial^4\,)\,+\,\frac{i}{4\,\sqrt 2\,\parp}\,(\,d^1\,{\bar d}_1\,+\,d^2\,{\bar d}_2\,-\,{\bar d}_1\,d^1\,-\,{\bar d}_2\,d^2\,) \\
&-\,\frac{i}{4\,\sqrt 2\,\parp}\,(\,d^3\,{\bar d}_3\,+\,d^4\,{\bar d}_4\,-\,{\bar d}_3\,d^3\,-\,{\bar d}_4\,d^4\,)\ .
\end{split}
\eea

\vskip 0.2cm

\noindent The four generators of the coset space $SO(4)/(SO(2)\,\times\,SO(2))$ are

\bea
\begin{split}
J^{12}\,=&\,x\,\partial^{12}\,-\,x^{12}\,\partial\,+\,\frac{i}{\sqrt 2}\,\parp\,\theta^1\,\theta^2\,-\,\frac{i\,\sqrt 2}{\parp}\,\partial^1\,\partial^2\,+\,\frac{i}{\sqrt 2\,\parp}\,d^1\,d^2\ , \\
{\bar J}_{12}\,=&\,{\bar x}\,{\bar \partial}_{12}\,-\,{\bar x}_{12}\,{\bar \partial}\,+\,\frac{i}{\sqrt 2}\,\parp\,{\bar \theta}_1\,{\bar \theta}_2\,-\,\frac{i\,\sqrt 2}{\parp}\,{\bar \partial}_1\,{\bar \partial}_2\,+\,\frac{i}{\sqrt 2\,\parp}\,{\bar d}_1\,{\bar d}_2\ , \\
J^{34}\,=&\,x\,\partial^{34}\,-\,x^{34}\,\partial\,+\,\frac{i}{\sqrt 2}\,\parp\,\theta^3\,\theta^4\,-\,\frac{i\,\sqrt 2}{\parp}\,\partial^3\,\partial^4\,+\,\frac{i}{\sqrt 2\,\parp}\,d^3\,d^4\ , \\
{\bar J}_{34}\,=&\,{\bar x}\,{\bar \partial}_{34}\,-\,{\bar x}_{34}\,{\bar \partial}\,+\,\frac{i}{\sqrt 2}\,\parp\,{\bar \theta}_3\,{\bar \theta}_4\,-\,\frac{i\,\sqrt 2}{\parp}\,{\bar \partial}_3\,{\bar \partial}_4\,+\,\frac{i}{\sqrt 2\,\parp}\,{\bar d}_3\,{\bar d}_4\ .
\end{split}
\eea

\noindent They satisfy the commutation relations

\bea
\begin{split}
\,[\,J^{12}\,,\,{\bar J}_{12}\,]\,&=\,-j\,-\,{\cal J}\ , \\
\,[\,J^{12}\,,J^{34}\,]\,&=\,[\,J^{12}\,,\,{\bar J}_{34}\,]\,=\,0\ , \\
\end{split}
\eea

\noindent and 

\bea
\begin{split}
\,[\,J^{34}\,,\,{\bar J}_{34}\,]\,&=\,-j\,+\,{\cal J}\ , \\
\,[\,J^{34}\,,J^{12}\,]\,&=\,[\,J^{34}\,,\,{\bar J}_{12}\,]\,=\,0\ . \\
\end{split}
\eea

\noindent The remaining generators are fairly straightforward to write down. The new ``plus" rotations read

\bea
J^{+\,12}&=&i\,x^{12}\,\parp\ ; \qquad \bar J^+_{~~12}~=~i\,\bar x_{12}\,\parp\ . 
\eea

\noindent The dynamical boosts are 
\bea
J^-&=&i\,x\,\frac{\partial\bar\partial\,+\,\frac {1}{2}\,\partial^{12}\,{\bar \partial}_{12}\,+\,\frac {1}{2}\,\partial^{34}\,{\bar \partial}_{34}\,}{\parp}\,-\,i\,x^-\,\partial\,+\,i\,\frac {\partial}{\parp}\,\Big\{\,\theta^m\,{\bar \partial}_m\,+\,\frac{i}{4\sqrt{2}\,\parp}\,(d^m\,{\bar d}_m\,-\,{\bar d}_m\,d^m)\,\Big\} \nonumber \\
&&-\,{\frac {1}{2}}\,\frac {{\bar \partial}_{12}}{\parp}\,{\biggl \{}\,\frac{1}{\sqrt 2}\,{\parp}\;\theta^1\,\,\theta^2\,-\,\frac{\sqrt 2}{\parp}\,\partial^1\,\partial^2\,+\,\frac {1}{\sqrt 2\,\parp}\,d^1\,d^2\,{\biggr \}} \nonumber \\
&&-\,{\frac {1}{2}}\,\frac {{\bar \partial}_{34}}{\parp}\,{\biggl \{}\,\frac{1}{\sqrt 2}\,{\parp}\;\theta^3\,\,\theta^4\,-\,\frac{\sqrt 2}{\parp}\,\partial^3\,\partial^4\,+\,\frac {1}{\sqrt 2\,\parp}\,d^3\,d^4\,{\biggr \}}\ ,
\eea

\noindent and its complex conjugate 

\bea
{\bar J}^-&=&i\,{\bar x}\,\frac{\partial\bar\partial\,+\,\frac {1}{2}\,\partial^{12}\,{\bar \partial}_{12}\,+\,\frac {1}{2}\,\partial^{34}\,{\bar \partial}_{34}\,}{\parp}~-~i\,x^-\,{\bar \partial}\,+\,i\,\frac {\partial}{\parp}\,\Big\{\,\theta^m\,{\bar\partial}_m\,+\,\frac{i}{4\sqrt{2}\,\parp}\,(\,d^m\,{\bar d}_m\,-\,{\bar d}_m\,d^m\,)\,\Big\} \nonumber \\
&&-\,{\frac {1}{2}}\,\frac {\partial^{12}}{\parp}\,{\biggl \{}\,\frac{1}{\sqrt 2}\,{\parp}\;{\bar \theta}_1\,\,{\bar \theta}_2\,-\,\frac{\sqrt 2}{\parp}\,{\bar \partial}_1\,{\bar \partial}_2\,+\,\frac {1}{\sqrt 2\,\parp}\,{\bar d}_1\,{\bar d}_2\,{\biggr \}} \nonumber \\
&&-\,{\frac {1}{2}}\,\frac {\partial^{34}}{\parp}\,{\biggl \{}\,\frac{1}{\sqrt 2}\,{\parp}\;{\bar \theta}_3\,\,{\bar \theta}_4\,-\,\frac{\sqrt 2}{\parp}\,{\bar \partial}_3\,{\bar \partial}_4\,+\,\frac {1}{\sqrt 2\,\parp}\,{\bar d}_3\,{\bar d}_4\,{\biggr \}}\ .
\eea

\vskip 0.2cm

\noindent In addition, we have new dynamical boosts obtained using the coset generators

\bea
\begin{split}
J^{-\,12}\,&\,=\,[\,J^-\,,\,J^{12}\,]\ ;\qquad \bar J^-_{12}\,=\,[\,\bar J^-\,,\,\bar J_{12}\,]\ ,\\
J^{-\,34}\,&\,=\,[\,J^-\,,\,J^{34}\,]\ ;\qquad \bar J^-_{34}\,=\,[\,\bar J^-\,,\,\bar J_{34}\,]\ ,
\end{split}
\eea

\noindent (which are not explicitly shown here). The dynamical supersymmetries in six dimensions are simply obtained by boosting the kinematical supersymmetries.

\bea
i\,[\,{\bar J}^-\,,\,{q^m}_+\,]\,\equiv\,{\cal Q}^m\ ;\qquad i\,[\,J^-\,,\,{\bar q}_{+\,m}\,]\,\equiv\,{\bar {\cal Q}}_m\ .
\eea

\noindent For example, the dynamical supersymmetries carrying a ``$1$" index read

\bea
{\cal Q}^1\,=\,\frac{\bar \partial}{\parp}\,q^1_+\,+\,\frac{\partial^{12}}{\parp}\,{\bar q}_{+\,2}\ ;\qquad {\bar {\cal Q}}_1\,=\,\frac{\partial}{\parp}\,{\bar q}_{+\,1}\,+\,\frac{{\bar \partial}_{12}}{\parp}\,q^2_+\ ,
\eea

\noindent and satisfy

\bea
\,\{\,{\cal Q}^1\,,\,{\bar {\cal Q}}_1\,\}\,=\,i\,{\sqrt 2}\,\frac{1}{\parp}\,(\,\partial\,{\bar \partial}\,+\,{\bar \partial}_{12}\,\partial^{12}\,)\ ,
\eea

\noindent permitting the introduction of central charges into the theory by setting $\partial^{12}$ to a constant, $Z^{12}$.

\vskip 1cm

\section{Oxidation From $d=4$ To $d=6$}

\vskip 0.5cm

\noindent Having built the six-dimensional SuperPoincar\'e algebra, we now focus on obtaining an invariant action to describe the $\mathcal N=(1,1)$ theory. 

\vskip 0.3cm
\noindent As outlined in our proposal, we permit the superfields, dependence on the two new directions
\bea
\phi\,=\,\phi\,(\,x^+,x^-,x,{\bar x},x^{12},{\bar x}_{12},x^{34},{\bar x}_{34}\,)\ ,
\eea

\noindent and define the extended six-dimensional d'Alembertian
\bea
\Box_6\,=\,2\,\partial\,{\bar \partial}\,+\,\partial^{12}\,{\bar \partial}_{12}\,+\,\partial^{34}\,{\bar \partial}_{34}\,-\,2\,\parp\,\parm\ .
\eea

\vskip 0.3cm
\noindent The key step is the generalization of the transverse derivatives. We define
\bea
{\nbar \nabla}\;=\;{\bar \partial}\,+\,\sigma\,{\bar d}_1\,{\bar d}_2\,\frac {\partial^{12}}{\parp}\,+\,\sigma\,{\bar d}_3\,{\bar d}_4\,\frac {\partial^{34}}{\parp}\ .
\eea

\noindent where $\sigma$ is a parameter that will be determined based on invariance requirements. The conjugate derivative reads

\bea
\nabla\;=\;\partial\,+\,\sigma\,d^1\,d^2\,\frac{{\bar \partial}_{12}}{\parp}\,+\,\sigma\,d^3\,d^4\,\frac{{\bar \partial}_{34}}{\parp}\ .
\eea

\vskip 0.3cm
\noindent Our proposal for the $\mathcal N=(1,1)$ Yang-Mills action in six dimensions is then simply

\bea
\label{d=611}
72\,\int\,d^6x\int d^4\theta\,d^4 \bar \theta\,{\cal L}\ ,
\eea

\noindent where

\bea
{\cal L}&=&-\bar\phi\,\frac{\Box_6}{\partial^{+2}}\,\phi
~+\frac{4g}{3}\,f^{abc}_{}\,\Big(\frac{1}{\partial^+_{}}\,\bar\phi^a_{}\,\phi^b_{}\,{\nbar \nabla}\,\phi^c_{}+\frac{1}{\partial^+_{}}\,\phi^a_{}\,\bar\phi^b_{}\,\nabla\,\bar\phi^c_{}\Big)\cr
&&-g^2f^{abc}_{}\,f^{ade}_{}\Big(\,\frac{1}{\partial^+_{}}(\phi^b\,\partial^+\phi^c)\frac{1}{\partial^+_{}}\,(\bar \phi^d_{}\,\partial^+_{}\,\bar\phi^e)+\frac{1}{2}\,\phi^b_{}\bar\phi^c\,\phi^d_{}\,\bar\phi^e\Big)\ .
\eea

\noindent In the next section, we will explicitly show that this action is left invariant by the $d=6$ SuperPoincar\'e algebra.

\vskip 1cm

\subsection{Invariance of the Action}

\noindent We intend to prove that the action in equation (\ref{d=611}) is $SO(4)\,-\,$invariant (Lorentz invariance in six dimensions follows once little group invariance has been established).

\vskip 0.3cm

\noindent We start by noting that the kinetic term is trivially $SO(4)$-invariant thanks to the inclusion of the two new derivatives in the d'Alembertian. The quartic interactions are obviously invariant since they do not depend on the transverse derivatives. Hence we focus purely on the cubic vertex

\bea
\frac{4g}{3}\,f^{abc}\,\int d^{10}x\,\int\,d^4\theta\,d^4\bar\theta\,\Big(\frac{1}{\parp}\,\bar\phi^a\,\phi^b\,\nbar\nabla\,\phi^c\,+\,\frac{1}{\parp}\,\phi^a\,{\nbar \phi}^b\,\nabla\,{\nbar \phi}^c\,\Big)\ .
\eea

\noindent Since this term is manifestly invariant under each $SO(2)$, we need consider only the coset variations. These coset generators vary both the superfields and the generalized derivatives. For example,

\bea
\delta_{J^{12}}\,\phi\,\equiv\,{\nbar \omega}_{12}\,J^{12}\,\phi\,=\,i\,{\sqrt 2}\;{\nbar \omega}_{12}\,{\parp}\,{\theta^1}\,{\theta^2}\,\phi\ ,
\eea

\noindent where the chiral constraint has been used.

\bea
\delta_{J^{12}}\,\phi\,\equiv\,{\nbar \omega}_{12}\,J^{12}\,\phi\,=\,{\nbar \omega}_{12}\,{\biggl \{}\,\frac{i}{\sqrt 2}\,\parp\,\theta^1\,\theta^2\,-\,i\,\frac{\sqrt 2}{\parp}\,\partial^1\,\partial^2\,+\,\frac{i}{\sqrt 2\,\parp}\,d^1\,d^2\,{\biggr \}}\,\phi\ ,
\eea
\noindent and

\bea
\delta_{J^{12}}\,{\nbar \nabla}\,=\,{\nbar \omega}_{12}\,[\,J^{12}\,,\,{\nbar \nabla}\,]\,=\,{\nbar \omega}_{12}\,{\biggl (}\,-\,\partial^{12}\,+\,\sigma\,{\bar d}_3\,{\bar d}_4\,\frac{\partial}{\parp}\,{\biggr )}\ .
\eea
\noindent Invariance under $SO(4)$ is verified by doing  a $\delta_J$ variation  on the entire cubic vertex.

\vskip 0.5cm

\subsection{The Variation}

\vskip 0.3cm

\noindent The proposed three-point function reads

\bea
{\mathbf {T}}\;+\;{\mathbf {T^*}}\;=\;{{\mathit {f}}_{abc}}\,{\int}\;{1\over \parp}\;{{\nbar \phi}^a}\;\;{\phi^b}\;\;{\nbar \nabla}\;{\phi^c}\;\;\;\;\;+\;\;\;\;\;{{\mathit {f}}_{abc}}\,{\int}\;{1\over \parp}\;{\phi^a}\;\;{{\nbar \phi}^b}\;\;{\nabla}\;{{\nbar \phi}^c}\ .
\eea

\noindent There are four coset generators and the aim is to show that each of them leaves this three-point function invariant. We start with the coset generator $J^{12}$ (the details of this calculation are presented in Appendix {\bf {A}}):

\bea
\begin{split}
\delta_{J^{12}}\,(\,\mathbf {T}\,)\,=\,-\,&\,[\,1\,+\,i{\sqrt 2}\,\sigma\,]\,\int\,\frac {1}{\parp}\,{\nbar \phi}^a\,\phi^b\,\partial^{12}\,\phi^c \\
\,+\,&\,\sigma\,\int\,\frac {1}{\parp}\,{\nbar \phi}^a\,\phi^b\,{\bar d}_3\,{\bar d}_4\,\frac{\partial}{\parp}\,\phi^c\ ,
\end{split}
\eea

\noindent and

\bea
\begin{split}
\delta_{J^{12}}\,(\,\mathbf {T^*}\,)\,=\,&\,[\,-\,\frac{i}{\sqrt 2}\,+\,\sigma\,]\,\int\,\frac{1}{\parp}\,\phi^a\,{\nbar \phi}^b\,d^1\,d^2\,\frac{\partial}{\parp}\,{\nbar \phi}^c \\
&\,i\,{\sqrt 2}\,\sigma\,\int\,\phi^a\,\frac{1}{\parp}\,{\nbar \phi}^b\,\partial^{12}\,\phi^c\ .
\end{split}
\eea

\noindent Choosing
\bea
\sigma~=~\frac{i}{2\,\sqrt 2}\ ,
\eea
\noindent ensures that
\bea
\delta_{J^{12}}\,(\,{\mathbf {T}}\,+\,{\mathbf {T^*}}\,)\,=\,0\ .
\eea

\noindent Thus the generalized derivative reads
\bea
{\nbar \nabla}\;=\;{\bar \partial}\,+\,\frac{i}{2\,\sqrt 2}\,{\bar d}_1\,{\bar d}_2\,\frac {\partial^{12}}{\parp}\,+\,\frac{i}{2\,\sqrt 2}\,{\bar d}_3\,{\bar d}_4\,\frac {\partial^{34}}{\parp}\ .
\eea

\noindent In deriving the above results, use has also been made of the inside-out relations, the identities listed in Appendix {\bf {B}} and numerous partial integrations with respect to $\parp$, $\bar \partial$ and $\partial$.

\vskip 0.2cm

\noindent Having fixed $\sigma$ we move to the other generators of the coset. Complex conjugation tells us that

\bea
\delta_{{\nbar J}_{12}}&\,(\,{\mathbf {T^*}}\,+\,{\mathbf {T}}\,)\,=\,0\ ,
\eea

\noindent while the remaining two variations proceed along identical lines

\bea
\delta_{J^{34}}\,(\,{\mathbf {T}}\,+\,{\mathbf {T^*}}\,)\,=\,0\ ;\qquad \delta_{{\nbar J}_{34}}\,(\,{\mathbf {T}}\,+\,{\mathbf {T^*}}\,)\,=\,0\ .
\eea

\noindent This completes the proof of $SO(4)$ invariance for the three-point function.

\vskip 0.3cm
\noindent Lorentz invariance in six dimensions is a direct consequence of little group invariance. Thus the $\mathcal N=(1,1)$ SuperYang-Mills theory in six dimensions is described by the light-cone action

\be
\label{zen}
\int d^6x\int d^4\theta\,d^4\bar \theta\,{\cal L}\ ,
\ee

\noindent where,

\bea
\begin{split}
{\cal L}&=&-\bar\phi^a\,\frac{\Box_6}{\partial^{+2}}\,\phi^a
~+\frac{4g}{3}\,f^{abc}_{}\,\Big(\frac{1}{\parp}\,\nbar\phi^a\,\phi^b\,\nbar\nabla\,\phi^c+\frac{1}{\parp}\,\phi^a\,\nbar\phi^b\,\nabla\,\nbar\phi^c\Big)\cr
&&-g^2f^{abc}_{}\,f^{ade}_{}\Big(\,\frac{1}{\parp}\,(\phi^b\,\parp\phi^c)\frac{1}{\parp}\,(\nbar \phi^d\,\parp\,\nbar\phi^e)+\frac{1}{2}\,\phi^b\nbar\phi^c\,\phi^d\,\nbar\phi^e\Big)\ .
\end{split}
\eea

\vskip 1cm

\section{$(\mathcal N=1,d=6)$ Yang-Mills through Truncation}

\noindent Oxidation (using the generalized derivatives) thus allows us to ``lift" a supersymmetric theory to its parent version. This procedure is however, supercharge-preserving and does not permit us access to theories with less supersymmetry. This is where supersymmetric truncation is useful.

\vskip 0.3cm

\noindent Supersymmetric truncation reduces the supersymmetries in a theory, one step at a time. We start by noting that~\cite{BT}
\bea
\label{trunc1}
\int\,{d^6}x\,{d^4}\theta\,{d^4}{\bar \theta}\,{\cal L}\;=\;\frac{1}{16}\,\int\,{d^6}x\,{d^3}\theta\,{d^3}{\bar \theta}\,{\bar d}_4\,d^4\,{\cal L}\,|_{\theta^4\,=\,{\bar \theta}_4\,=0}\ .
\eea
\noindent This truncation, produces two kinds of superfields, the first being bosonic
\bea
\label{trunf}
\phi^{(3)}\,=\,\phi\,|_{\theta^4\,=\,{\bar \theta}_4\,=0}\ ,
\eea
\noindent and the second fermionic
\bea
\psi_4\,=\,{\bar d}_4\,\phi\,|_{\theta^4\,=\,{\bar \theta}_4\,=0}\ .
\eea
\noindent This fermionic superfield may be eliminated in favor of the bosonic one thanks to the `inside-out" constraints in equation (\ref {ioc}). Additional truncation involves rewriting
\bea
\label{trunc2}
\int\,{d^6}x\,{d^3}\theta\,{d^3}{\bar \theta}\,{\cal L}\;=\;-\,\frac{1}{9}\,\int\,{d^6}x\,{d^2}\theta\,{d^2}{\bar \theta}\,{\bar d}_3\,d^3\,{\cal L}\,|_{\theta^3\,=\,{\bar \theta}_3\,=0}\ ,
\eea
which generates two new and {\it {independent}} superfields,
\bea
\phi^{(2)}\,=\,\phi^{(3)}\,|_{\theta^3\,=\,{\bar \theta}_3\,=0}\ ,
\eea
\noindent and
\bea
\psi_3\,=\,{\bar d}_3\,\phi^{(3)}\,|_{\theta^3\,=\,{\bar \theta}_3\,=0}\ ,
\eea
\noindent We set the fermionic superfield (which produces Wess-Zumino couplings~\cite{BT}) to zero and focus exclusively on the bosonic superfield. 

\vskip 0.2cm

\noindent The doubly truncated bosonic superfield now reads
\bea
\phi^{(2)}\,(y)\,=\,\frac{1}{\parp}\,A\,(y)\,+\,i\,{\sqrt 2}\;\theta^1\,\theta^2\,{\nbar C}_{12}\,(y)\,+\,\frac{i}{\parp}\,\theta^1\,{\bar \chi}_1(y)\,+\,\frac{i}{\parp}\,\theta^2\,{\bar \chi}_2(y)\ ,
\eea
\noindent and carries the degrees of freedom relevant to the $N=1$ theory in six-dimensions (and the $N=2$ theory in four dimensions, an issue we will return to shortly). 

\vskip 0.5cm

\noindent We now apply relations (\ref{trunc1}) and (\ref{trunc2}) to our six-dimensional $\mathcal N=(1,1)$ action (equation \ref{zen}) to obtain the light-cone superspace description of $(\mathcal N=1,d=6)$ Yang-Mills. The calculation is fairly straightforward and yields

\be
\label{six}
\int\,d^6x\int d^2\theta\,d^2\bar \theta\,{\cal L}\ ,
\ee

\noindent where,

\bea
\label{six1}
\begin{split}
{\cal L}=&-\,2\,{\bar \phi}^{(2)\,a}\,{\Box}\,\phi^{(2)\,a}\,{\mbox {\hskip 6cm}} \\
&+\,4\,g\,f^{abc}\,{\biggl \{}\,\,\parp\,\phi^{(2)\,a}\,{\nbar \phi}^{(2)\,b}\,{\bar \partial}\,\phi^{(2)\,c}\,+\,\frac{i}{2\,\sqrt 2}\,\parp\,{\nbar \phi}^{(2)\,a}\,\phi^{(2)\,b}\,{\bar d}_1\,{\bar d}_2\,\frac{\partial^{12}}{\parp}\,\phi^{(2)\,c} \\
&+\,\frac{i}{2\,\sqrt 2}\,\parp\,\phi^{(2)\,a}\,{\nbar \phi}^{(2)\,b}\,d^1\,d^2\,\frac{{\bar \partial}_{12}}{\parp}\,{\nbar \phi}^{(2)\,c}\,{\biggr \}}\,+\,4\,g\,f^{abc}\,{\biggl \{}\,{\mbox {complex conjugate}}\,{\biggr \}} \\
&-\,\frac{g^2}{2}\,f^{abc}\,f^{ade}\frac{d^2}{\parp}\,(\parp\,\phi^{(2)\,b}\,{\bar \phi}^{(2)\,c})\,\frac{{\bar d}^2}{\parp}\,(\parp\,{\bar \phi}^{(2)\,d}\,\phi^{(2)\,e})\ .{\mbox {\hskip 1cm}}
\end{split}
\eea

\vskip 0.2cm

\noindent We note that the superfields in the above expression are no longer constrained (although they still satisfy the chirality relations).

\vskip 0.5cm

\noindent As expected, this action can be reduced to four dimensions, producing the $(\mathcal N=2,d=4)$ theory. This is easily verified - we simply remove the superfield dependence on the new coordinates thus setting
\bea
\partial^{12}\,=\,{\bar \partial}_{12}\,\rightarrow\,0\ .
\eea

\noindent This results in the light-cone description of four-dimensional $\mathcal N=2$ Yang-Mills~\cite{JT,GPK}

\be
\int d^4x\int d^2\theta\,d^2\bar \theta\,{\cal L}\ ,
\ee

\bea
\label{four2}
\begin{split}
{\cal L}=&-\,2\,\bar\phi^{(2)\,a}\,{\Box}\,\phi^{(2)\,a}{\mbox {\hskip 5cm}} \\
&+\,\frac{4}{3}\,g\,f^{abc}\,{\biggl \{}\,\parp\,\phi^{(2)\,a}\,{\nbar \phi}^{(2)\,b}\,{\bar \partial}\,\phi^{(2)\,c}\,+\,\parp\,{\nbar \phi}^{(2)\,a}\,\phi^{(2)\,b}\,\partial\,{\nbar \phi}^{(2)\,c}\,{\biggr \}} \\
&-\,\frac{g^2}{2}\,f^{abc}\,f^{ade}\frac{d^2}{\parp}\,(\parp\,\phi^{(2)\,b}\,{\bar \phi}^{(2)\,c})\,\frac{{\bar d}^2}{\parp}\,(\parp\,\bar \phi^{(2)\,d}\,\phi^{(2)\,e})\ .{\mbox {\hskip 1cm}}
\end{split}
\eea

\vskip 1cm

\section{Concluding Remarks}

\vskip 0.5cm

\noindent Light-cone superspace offers an excellent stage to build Lorentz-invariant interactions of massless particles with arbitrary helicities. In this language for example, the entire classical $PSU(2,2|4)$-invariant $(\mathcal N=4,d=4)$ action can be written as the square of a single fermionic superfield~\cite{ABR3}. The techniques presented here (and in the quoted references) should prove extremely useful in building light-cone actions for theories whose form is still unknown.

\vskip 1cm

\noindent {\bf {Acknowledgments}}

\vskip 0.3cm

\noindent It is a pleasure to thank Professors Pierre Ramond and Lars Brink for helpful discussions.

\newpage

\renewcommand{\theequation}{A-\arabic{equation}}
  \setcounter{equation}{0}  
  \section*{Appendix {\bf {A}}}  

\vskip 0.3cm

\subsection*{Variation: $\delta_{J^{12}}\,(\,\mathbf {T}\,)$}

\vskip 0.2cm

\noindent From varying the first superfield, 
\bea
\begin{split}
\delta_{J^{12}}\,(\,\frac{1}{\parp}\,{\nbar \phi}^a\,)\,\phi^b\,{\nbar \nabla}\,\phi^c\,&\,=\,\frac{i}{\sqrt 2}\,\theta^1\,\theta^2\,{\biggl \{}\,{\nbar \phi}^a\,\phi^b\,{\nbar \nabla}\,\phi^c\,+\,\frac{1}{{\parp}^2}\,{\nbar \phi}^a\,{{\parp}^2}\,\phi^b\,{\nbar \nabla}\,\phi^c \nonumber \\
&\,+\,2\,\frac{1}{{\parp}^2}\,{\nbar \phi}^a\,\parp\,\phi^b\,\parp\,{\nbar \nabla}\,\phi^c\,+\,\frac{1}{{\parp}^2}\,{\nbar \phi}^a\,\phi^b\,{{\parp}^2}\,{\nbar \nabla}\,\phi^c\,{\biggr \}} \\
&\,+\,i\,{\sqrt 2}\,\sigma\,\theta^1\,(\,\frac{1}{{\parp}^2}\,{\nbar \phi}^a\,\parp\,\phi^b\,{\bar d}_1\,\partial^{12}\,\phi^c\,+\,\frac{1}{{\parp}^2}\,{\nbar \phi}^a\,\phi^b\,{\bar d}_1\,\partial^{12}\,\parp\,\phi^c\,) \\
&\,+\,i\,{\sqrt 2}\,\sigma\,\theta^2\,(\,\frac{1}{{\parp}^2}\,{\nbar \phi}^a\,\parp\,\phi^b\,{\bar d}_2\,\partial^{12}\,\phi^c\,+\,\frac{1}{{\parp}^2}\,{\nbar \phi}^a\,\phi^b\,{\bar d}_2\,\partial^{12}\,\parp\,\phi^c\,) \ .
\end{split}
\eea
\noindent The variation on the second superfield is

\bea
\frac{1}{\parp}\,{\nbar \phi}^a\,(\,\delta_{J^{12}}\,\phi^b\,)\,{\nbar \nabla}\,\phi^c\,=\,i\,{\sqrt 2}\,\theta^1\,\theta^2\,\frac{1}{\parp}\,{\nbar \phi}^a\,\parp\,\phi^b\,{\nbar \nabla}\,\phi^c\ ,
\eea
\noindent The third contribution is from the newly introduced derivative and reads

\bea
\frac{1}{\parp}\,{\nbar \phi}^a\,\phi^b\,(\,\delta_J\,{\nbar \nabla}\,)\,\phi^c\,=\,-\,\frac{1}{\parp}\,{\nbar \phi}^a\,\phi^b\,\partial^{12}\,\phi^c\,+\,\sigma\,\frac{1}{\parp}\,{\nbar \phi}^a\,\phi^b\,{\bar d}_3\,{\bar d}_4\,\frac{\partial}{\parp}\,\phi^c\ .
\eea
\noindent The final term in the variation is

\bea
\begin{split}
\frac{1}{\parp}\,{\nbar \phi}^a\,\phi^b\,{\nbar \nabla}\,(\,\delta_{J^{12}}\,\phi^c\,)\,&=\,-\,i{\sqrt 2}\,\sigma\,\frac{1}{\parp}\,{\nbar \phi}^a\,\phi^b\,\partial^{12}\,\phi^c\,+\,i\,{\sqrt 2}\,\sigma\,\theta^1\,\frac{1}{\parp}\,{\nbar \phi}^a\,\phi^b\,{\bar d}_1\,\partial^{12}\,\phi^c \\
&+\,i\,{\sqrt 2}\,\sigma\,\theta^2\,\frac{1}{\parp}\,{\nbar \phi}^a\,\phi^b\,{\bar d}_2\,\partial^{12}\,\phi^c \ .
\end{split}
\eea
\noindent These terms simplify greatly (after some partial integrations) to

\bea
\delta_{J^{12}}\,(\,\int\,\frac {1}{\parp}\,{\nbar \phi}^a\,\phi^b\,{\nbar \nabla}\,\phi^c\,)\,=\,-\,[\,1\,+\,i{\sqrt 2}\,\sigma\,]\,\int\,\frac {1}{\parp}\,{\nbar \phi}^a\,\phi^b\,\partial^{12}\,\phi^c\,+\,\sigma\,\int\,\frac {1}{\parp}\,{\nbar \phi}^a\,\phi^b\,{\bar d}_3\,{\bar d}_4\,\frac{\partial}{\parp}\,\phi^c\ .
\eea

\vskip 1cm

\subsection*{Variation: $\delta_{J^{12}}\,(\mathbf {T^*}\,)$}

\bea
\delta_{J^{12}}\,(\,\frac{1}{\parp}\,\phi^a\,)\,{\nbar \phi}^b\,\nabla\,{\nbar \phi}^c\,&\,=\,i\,{\sqrt 2}\,\theta^1\,\theta^2\,\phi^a\,{\nbar \phi}^b\,\nabla\,{\nbar \phi}^c\,
\eea

\bea
\begin{split}
\frac{1}{\parp}\,\phi^a\,(\,\delta_{J^{12}}\,{\nbar \phi}^b\,)\,\nabla\,{\nbar \phi}^c\,=\,&\frac{i}{\sqrt 2}\,\theta^1\,\theta^2\,\frac{1}{\parp}\,\phi^a\,\parp\,{\nbar \phi}^b\,\nabla\,{\nbar \phi}^c\,-\,\frac{i}{\sqrt 2}\,\frac{1}{\parp}\,\phi^a\,\frac{\partial^1\,\partial^2}{\parp}\,{\nbar \phi}^b\,\nabla\,{\nbar \phi}^c \\
+\,&\frac{i}{\sqrt 2}\,\frac{1}{\parp}\,\phi^a\,\frac{d^1\,d^2}{\parp}\,{\nbar \phi}^b\,\nabla\,{\nbar \phi}^c\ .
\end{split}
\eea

\bea
\frac{1}{\parp}\,\phi^a\,{\nbar \phi}^b\,(\,\delta_{J^{12}}\,\nabla\,)\,{\nbar \phi}^c\,=\,\sigma\,\frac{1}{\parp}\,\phi^a\,{\nbar \phi}^b\,d^1\,d^2\,\frac{\partial}{\parp}\,{\nbar \phi}^c\ .
\eea

\bea
\begin{split}
\frac{1}{\parp}\,\phi^a\,{\nbar \phi}^b\,\nabla\,(\,\delta_{J^{12}}\,{\nbar \phi}^c\,)\,=\,&\frac{i}{\sqrt 2}\,\theta^1\,\theta^2\,\frac{1}{\parp}\,\phi^a\,{\nbar \phi}^b\,\nabla\,\parp\,{\nbar \phi}^c\,-\,i\,\sqrt 2\,\frac{1}{\parp}\,\phi^a\,{\nbar \phi}^b\,\nabla\,\frac{\partial^1\,\partial^2}{\parp}\,{\nbar \phi}^c \\
+\,&\frac{i}{\sqrt 2}\,\frac{1}{\parp}\,\phi^a\,{\nbar \phi}^b\,\nabla\,\frac{d^1\,d^2}{\parp}\,{\nbar \phi}^c\ .
\end{split}
\eea

\vskip 2cm

\renewcommand{\theequation}{B-\arabic{equation}}
  \setcounter{equation}{0}  
  \section*{Appendix {\bf {B}}}  

\vskip 0.3cm

\subsection*{Useful Identities}

\vskip 0.2cm

The inside-out constraints read

\bea
{{\bar d}_p}\,{{\bar d}_q}\,{\phi}\,=\,{\frac {1}{2}}\,{\epsilon_{pqmn}}\,{d^m}\,{d^n}\,{\nbar \phi}\;\;\;\;\;\;\;\;\;\;;\;\;\;\;\;\;\;\;\;\;{{\bar d}_p}\,{{\bar d}_q}\,{{\bar d}_m}\,{{\bar d}_n}\,{\phi}\,=\,2\,{\epsilon_{pqmn}}\,{{\parp}^2}\,{\nbar \phi}
\eea

\vskip 0.3cm

\noindent Consequence \#1,

\bea
\label{con1}
{{\mathit {f}}_{abc}}\,{\int}\;{\frac {1}{{\parp}^2}}\,{{\nbar \phi}^a}\;{\phi^b}\;{\bar \partial}\,{\phi^c}\;=\;0
\eea

\vskip 0.2cm
\noindent Proof:
\vskip 0.2cm

\bea
\begin{split}
&{\int}\;{\frac {1}{{\parp}^2}}\,{{\nbar \phi}^a}\;{\phi^b}\;{\bar \partial}\,{\phi^c}={\int}\;{\frac {1}{{\parp}^2}}\,{{\nbar \phi}^a}\;{\frac {3\,{d^4}}{{\parp}^2}}\,{{\nbar \phi}^b}\;{\bar \partial}\,{\phi^c}\,=\,{\int}\;{\frac {3\,{d^4}}{{\parp}^2}}\,{{\nbar \phi}^a}\;{\frac {1}{{\parp}^2}}\,{{\nbar \phi}^b}\;{\bar \partial}\,{\phi^c} \\
=&{\int}\;\,{\phi^a}\;{\frac {1}{{\parp}^2}}\,{{\nbar \phi}^b}\;{\bar \partial}\,{\phi^c}\,=\,0\;\;{\mbox {(due to symmetry between the $a$ and $b$ indices)}}
\end{split}
\eea

\vskip 0.5cm

\noindent Consequence \#2

\bea
\label{con2}
{{\mathit {f}}_{abc}}\,{\int}\;{\frac {1}{\parp}}\,{{\nbar \phi}^a}\;{\frac {1}{\parp}}\,{\phi^b}\;{d^m}\,{d^n}\,{\partial}\,{{\nbar \phi}^c}\;=\;0
\eea

\vskip 0.2cm
\noindent Proof:
\vskip 0.2cm

\bea
\begin{split}
{\int}\;{\frac {1}{\parp}}\,{{\nbar \phi}^a}\;{\frac {1}{\parp}}\,{\phi^b}\;{d^m}\,{d^n}\,{\partial}\,{{\nbar \phi}^c}\;=&\;{\int}\;{\frac {1}{\parp}}\,{d^m}\,{d^n}\,{{\nbar \phi}^a}\;{\frac {1}{\parp}}\,{\phi^b}\;{\partial}\,{{\nbar \phi}^c} \\
=&\,{\frac {1}{2}}\,{\epsilon^{mnpq}}\,{\int}\;{\frac {1}{\parp}}\,{{\bar d}_p}\,{{\bar d}_q}\,{\phi^a}\;{\frac {1}{\parp}}\,{\phi^b}\;{\partial}\,{{\nbar \phi}^c} \\
=&\,{\frac {1}{2}}\,{\epsilon^{mnpq}}\,\;{\int}\;{\frac {1}{\parp}}\,{\phi^a}\;{\frac {1}{\parp}}\,{{\bar d}_p}\,{{\bar d}_q}\,{\phi^b}\;{\partial}\,{{\nbar \phi}^c}\ , \\
=&\,0\;\;\;{\mbox {($a-b$ symmetry)}}
\end{split}
\eea

\end{document}